\documentclass[a4paper,11pt]{article}
\usepackage{aaskaiid}
\usepackage{orcidlink}

\title{Commensal image plane transient search methods with the SKAO}
\ShortTitle{Commensal transient search methods}
\ShortName{The Transients Science Working Group} 

\author[1]{Alex Andersson\orcidlink{0000-0003-2734-1895}}
\author[2,3]{Dougal Dobie\orcidlink{0000-0003-0699-7019}}
\author[4]{Natasha Hurley-Walker\orcidlink{0000-0002-5119-4808}}
\author[5]{Hao Qiu\orcidlink{0000-0002-9586-7904}}
\author[1]{Kaustubh Rajwade\orcidlink{0000-0002-8043-6909}}
\author[\dagger,6,7]{Antonia Rowlinson\orcidlink{0000-0002-1195-7022}}
\author[\dagger,2,3]{Iris de Ruiter\orcidlink{0000-0002-4752-5467}}
\author[8]{Fabian Sch\"ussler\orcidlink{0000-0003-1500-6571}}
\author[9,10,11]{Oleg M. Smirnov\orcidlink{0000-0003-1680-7936}}
\author[12]{Ben Stappers\orcidlink{0000-0001-9242-7041}}
\author[]{The Transients Science Working Group}

\affiliation[\dagger]{Chapter co-ordinator}
\affiliation[1]{Astrophysics, Department of Physics, University of Oxford, Denys Wilkinson Building, Keble Road, Oxford OX1 3RH, UK}
\affiliation[2]{Sydney Institute for Astronomy, School of Physics, The University of Sydney, NSW 2006, Australia}
\affiliation[3]{ARC Centre of Excellence for Gravitational Wave Discovery (OzGrav), Hawthorn, VIC 3122, Australia}
\affiliation[4]{International Centre for Radio Astronomy Research, Curtin University, Kent St, Bentley WA 6102, Australia}
\affiliation[5]{SKA Observatory, 26 Dick Perry Avenue, Kensington WA 6151, Australia}
\affiliation[6]{Anton Pannekoek Institute for Astronomy, University of Amsterdam, Science Park 904, P.O. Box 94249, 1090GE Amsterdam, The Netherlands}
\affiliation[7]{ASTRON, the Netherlands Institute for Radio Astronomy, Postbus 2, NL-7990 AA Dwingeloo, The Netherlands}
\affiliation[8]{IRFU, CEA, Université Paris-Saclay, F-91191 Gif-sur-Yvette, France}
\affiliation[9]{Centre for Radio Astronomy Techniques \& Technologies (RATT), Department of Physics and Electronics, Rhodes University, Makhanda 6139, South Africa}
\affiliation[10]{South African Radio Astronomy Observatory (SARAO), Cape Town 7700, South Africa}
\affiliation[11]{Institute for Radioastronomy, National Institute of Astrophysics (INAF IRA), Bologna 40129, Italy}
\affiliation[12]{ Jodrell Bank Centre for Astrophysics, Department of Physics and Astronomy, The University of Manchester, Manchester M13 9PL, UK}
\emailAdd{b.a.rowlinson@uva.nl}
\emailAdd{iris.deruiter@sydney.edu.au}

\abstract{
This chapter outlines the key, state of the art, techniques required to conduct commensal image plane transient searches. Using significant experience and expertise from conducting transient searches with the SKA pathfinders, we have developed efficient fast imaging strategies, automated transient detection pipelines, determined careful filtering techniques to avoid artifacts, and identified the most useful triggered reprocessing tools. Using these strategies, we will be ideally placed to optimally and reliably detect transient sources from commensal image plane transient surveys. The tools and methods presented in this chapter can be used for all SKA-Mid and SKA-Low array deployments from AA$^*$ inclusive.
}

\begin{document}
\maketitle

\section{Introduction}

Astrophysical transient sources provide unique ways that we can probe the Universe as they typically involve the strongest gravitational forces, strongest magnetic fields, highest densities and largest relativistic effects. They enable us to probe fundamental physics in ways that are impossible to recreate on Earth. However, many of these events are also typically rare, requiring specialized techniques to find them. There are typically two ways we can study these events with radio telescopes, firstly follow-up of transients detected by other multi-messenger facilities and secondly by conducting large unbiased transient surveys. Unbiased transient surveys refers to surveys that are not targeting specific known transient populations. As large volumes of data are required to capture the rare transient events, these surveys are not from dedicated observations and instead typically use commensal datasets that are being observed for different science cases. We note that these commensal datasets may also have their own intrinsic biases, for example observations of the Galactic plane are naturally biased towards Galactic sources whereas observations of the most massive galaxy clusters will be biased towards extragalactic sources. Here, we focus on large unbiased transient surveys using commensally observed datasets.

When a new facility with unique capabilities starts up operations, this often leads to serendipitous and unexpected detections of transient sources. For instance, Gamma-Ray Bursts (GRBs) were discovered due to the launch of whole sky gamma-ray monitors \citep{klebesadel1973} and in optical the detection of `The Cow' in wide-field optical surveys led to the new class of transients known as Fast Blue Optical Transients \citep[FBOTs;][]{prentice2018}. Alternatively, the development of new survey techniques and increased computational resources can also lead to new discoveries, such as the detection of the first Fast Radio Bursts in archival Parkes data \citep{lorimer2007}.

In the radio image plane, synchrotron transient detection has been challenging. Many of the known standard synchrotron transients (such as GRBs, X-ray binaries and supernovae) are expected to be long duration and faint at radio frequencies, leading to the expectation that deep, wide-field surveys are needed with observation separations of days to years to find them in unbiased surveys \citep[e.g.][]{metzger2015}. SKAO will have the sensitivity to search for these sources, however dedicated unbiased surveys will cost significant observing time (see \cite{GemmaAnderson01.2026.SKA} and \cite{Colombo01.2026.SKA} for details regarding targeted follow-up of these events).

On shorter durations, seconds to $\lesssim$hours, teams using the SKA pathfinder facilities have conducted large surveys using their large fields of view and excellent snapshot imaging capabilities. They unexpectedly discovered a new population of Long Period Transients \citep[LPTs; e.g.][]{hurleywalker2022}. See \cite{Qiu01.2026.SKA} for more details regarding the LPTs in the era of SKAO and \cite{AlexAndersson01.2026.SKA} for prospects of detecting unknown transients in SKAO commensal surveys. However, to find more of these sources, very large volumes of data must be efficiently processed. SKAO will be an outstanding facility for finding these sources, but the volume of data required is likely prohibitive for dedicated surveys.

\begin{figure*}
    \centering
    \includegraphics[width=0.9\textwidth]{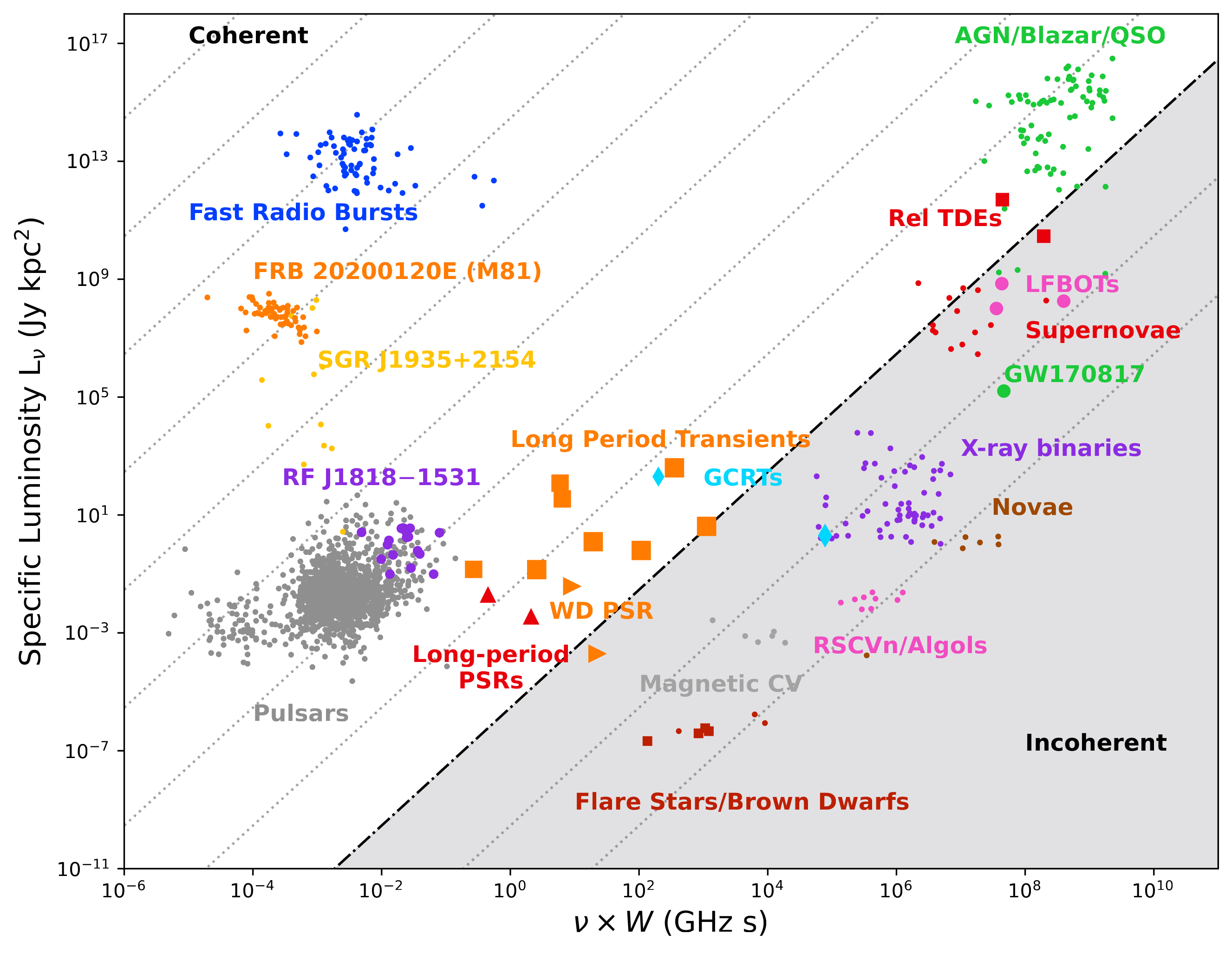}
    \protect \caption{ Transient phase space showing radio luminosity versus the product of timescale and observing frequency for different transient source classes, following \citet{cordes2004dynamic}.  Note that the luminosity assumes sources are beamed into only $1$\,sr and no relativistic beaming, which may or may not be appropriate for individual objects, while the timescales are just the observed variability timescales and ignore more constraining limits such as the finite sizes of e.g. stellar sources. For sources with relativistic beaming, the intrinsic luminosity of the source can be significantly lower than that observed leading to the true brightness temperature being significantly lower. Alternatively, for some stellar sources the intrinsic luminosity is significantly higher and the true brightness temperature could be significantly higher.  The diagonal lines show contours of brightness temperature, with coherent emitters having $T_B>10^{12}\,$K.  Adapted from \citet{pietka2015variability} and \citet{nimmo2022burst}. Figure from \protect \cite{murphy2025}, see their Figure 1 caption for full reference list.}
    \label{fig:phasespace}
\end{figure*}

In Figure \ref{fig:phasespace}, we show the known transient population at radio frequencies. The unbiased transient surveys in the image plane can target sources with transient durations $>$1 second. This search data does not need to be of particular dedicated fields, rather any data is sufficient. The teams using the SKA pathfinder facilities have demonstrated the significant value of commensal surveys \citep[e.g.][]{deruiter2024,wang2021askap,rowlinson2016}.

This chapter will consider how we can implement transient surveys with commensal survey data from the SKAO. There have been significant advances in optimising transient and variability searches with the SKA pathfinders on a range of timescales, that can be directly applied to our strategies for processing SKAO data. We will outline these advances, explain why they are implemented and describe outstanding issues with processing these data. These advances can be applied to observations from all deployments of SKA-Mid and SKA-Low starting from AA$^*$.

\section{Imaging strategies}

As outlined in the previous section, the radio transient phase space covers a wide range of timescales, from milliseconds to decades. Recent advances in processing techniques and detection pipelines have blurred the traditional division between short-duration transients that were best identified and studied with beamformed techniques, and long duration transients that were best detected and characterized with imaging techniques. 
Specifically, the use of model-subtracted snapshot images has allowed the detection of transients with durations of $\sim$seconds to minutes in the imaging domain. This technique enables us to detect interacting compact object binaries, long-period transients, radio stars \citep{Beri01.2026.SKA,Qiu01.2026.SKA,Driessen01.2026.SKA}, and rapidly scintillating sources \citep{wang2021askap}. Additionally, long-period pulsars, Rotating Radio Transients, and even FRBs \citep[e.g.][]{Curtin01.2026.SKA,Caleb02.2026.SKA} have been detected using short integration snapshots in the imaging domain \citep[see e.g.][]{caleb2022discovery,2025ApJ...981..143M,andrianjafy2023image}.

Alternatively, a more traditional approach takes two or more images, separated by some interval of time, and compares the sources detected in each of those images to identify differences between them. These observations are usually days, months or years apart, making them sensitive to the radio signatures of gamma-ray bursts, tidal disruption events, active galactic nuclei, supernovae, and novae. Below we will discuss the two approaches in detail. 

\subsection{Short snapshot surveys} \label{sec:short_snapshot_surveys}

In some optical transient surveys, such as the Legacy Survey of Space and Time (LSST), \textit{difference imaging}— subtracting consecutive snapshots or subtraction of a deep reference image — removes constant emission and reveals transients \citep[see e.g.][]{liu2024testing}. This approach is less effective in radio, where sparse $u,v$-coverage in short integrations produces irregular point spread functions and structured noise, generating many artefacts upon subtraction - unless the images are very close together in time (typically of order minutes due to the rotation of the point spread function due to the Earth's rotation). An alternative is to create model-subtracted snapshot images: subtract a deep integration sky model from the $u,v$-data before imaging. Images at various cadences then show only deviations from the sky model.\\

Model-subtracted methods have three main advantages over standard cleaned continuum imaging methods: (1) Imaging is faster, as primary beam correction and cleaning are unnecessary when the field is source-free. This is essential when imaging long observations at second-scale cadences. (2) Images contain only sources absent from, or variable relative to, the model, simplifying transient searches. (3) Subtracting the deep model can reduce confusion noise for some facilities, enabling deeper searches \citep[see e.g.][]{fijma2024new}. Sky model subtracted snapshots are therefore the ideal data product to feed into transient pipelines \citep[see e.g.][]{wang2023radio, de2024transient, horvath2025long, smirnov2025mining}. Once a transient source has been detected, additional images without model-subtraction can be made to show the source environment and obtain accurate flux density measurements and astrometry.\\

A commensal snapshot imaging pipeline for the SKAO would enable the exploration of large sky areas necessary to uncover the most exotic and rare types of transient sources. To make this search effective, the model-subtracted snapshot images should be created during the processing stage, as both a deep sky model and visibility data are required. Much progress has been made in recent years in the transient pipelines that process such images (see Section 3.1). In the SKAO-era a snaphshot fast imaging pipeline will run on the SKAO Central Processing Facilities, in near-real-time. The SKAO fast imaging pipeline will enter science verification in Cycle 1, with shared risk observations from Cycle 2.

\subsection{Long surveys (multiple observations)} \label{sec:long_surveys}

For longer durations, days or longer, commensal transient searches can use the standard continuum output images of individual observations. Some science cases will require multiple observations of the same field to reach the required sensitivity. These observations could be separated by days, months or even years, providing an opportunity to comensally search for long duration transients and variability. Additionally, there may be overlap between fields observed by different science cases giving additional epochs to search for transients and variables.

The different observations of each field will typically not be optimally separated in time, leading to sparse and unevenly sampled transient surveys. Using commensal observations from SKA Pathfinders, techniques have been optimised to enable analysis and interpretation of transient rates from these complex datasets \citep{chastain2022}.

The output required for these surveys is a deep, cleaned image of the field containing all the detectable sources per observation. In the SKAO era, this image will be a standard output for observations produced by the SKAO Central Processing Facilities. A transient pipeline, running automatically on the Central Processing Facility or offline using the SKAO Regional Centre Network (SRCNet), can search for transient and variable sources in these output images.

\section{Transient detection pipelines}

Due to the large volume of images that require processing for commensal transient searches, it is essential to use fully automated transient detection pipelines. In this section, we highlight the current state-of-the-art pipelines and strategies for extracting interesting sources.

\subsection{From images to light curves}

The key steps behind the majority of radio transient detection pipelines are the same, with key differences being in the implementation. Here we consider these underlying steps and then state key differences between the existing pipelines.

The typical inputs for any unbiased transient survey pipeline are the images, which are typically created by the observatory specific imaging pipelines. Sometimes calibration and imaging fails or the output image is of poor quality, thus the key first step to any transient detection pipeline is ensuring that the image quality is sufficient. Often a simple rms cut based on the standard image quality in the dataset is sufficient, e.g. by rejecting all images whose rms values are more than 3$\sigma$ deviant from the average rms in the dataset. 

Following image quality control, the next standard step is to search for sources in the images. Typically, any radio image plane source finder can be used but often these unbiased surveys want good point source flux density recovery and for the source finder to be optimized for processing speed. The transient pipeline then conducts source association, across time and frequency, with the previously detected source list to build source light curves and add new sources to the detected source list. These light curves are then typically stored in a database that can be queried to search for new sources or sources with significantly variable light curves. 

``Fast'' and ``standard'' pipelines can differ substantially in their approach, science applications and required infrastructure. Several ``fast'' pipelines ingest visibilities and carry out their own custom imaging which, while not as computationally expensive as creating science-quality deep continuum images, requires substantially more compute than ``standard'' pipelines which ingest existing images (and sometimes catalogues). The deployment of these pipelines will therefore differ substantially in the SKAO era -- ``fast'' pipelines will need to be integrated into the main SKAO science workflows with final data products supplied to users, while in most cases it will remain feasible for users to run ``standard'' pipelines (provided by themselves or the SKAO) on SRCNet. ``Fast'' pipelines are also generally deployed to find highly intermittent or time-sensitive sources, which in the SKAO era may necessitate the inclusion of some alert infrastructure into the pipelines, while ``standard'' pipelines search for slower evolving sources which do not require immediate user intervention.

\begin{itemize}
    \item TraP - The LOFAR Transients Pipeline \citep[{\sc TraP};][]{swinbank2015}\footnote{\url{https://github.com/transientskp/tkp}} is a publicly available pipeline that follows all of the fundamental steps outlined above and can handle any standard fits format image. TraP uses the {\sc Python Source Extractor} \citep[{\sc PySE};][]{carbone2018} which is significantly optimized for processing speed and for point source flux density recovery. The TraP is currently undergoing a significant upgrade to increase processing speed and to enable it to be fully incorporated in the automated data processing pipeline for LOFAR2.0\footnote{ \url{https://git.astron.nl/RD/trap}}. These improvements allow all LOFAR2.0 imaging observations to be automatically processed by TraP, producing a legacy, publicly queryable, database of source light curves.
    \item VAST - The Variables And Slow Transients Pipeline \citep[{\sc VAST};]{adam_stewart_2024_14048598} is an open-source transient detection pipeline optimised for observations conducted with ASKAP. It ingests the standard ASKAP data products (images and auxiliary maps produced with {\sc ASKAPsoft} \citep{cornwell2011askap}, source catalogues produced by {\sc Selavy} \citep{Whiting_Humphreys_2012}) and carries out source association, forced photometry and calculation of statistics. Data are queryable via a dedicated web interface, or programmatically using the {\sc vast-tools} Python package \citep{adam_stewart_2025_15363128}.
    \item VASTER - The VAST fast detection pipeline \citep[{\sc VASTER};][]{wang2021} is optimised for searching for minute-timescale variability in ASKAP observations on a per-beam basis. The pipeline generates a deep model that is then subtracted from the visibilities, and then forms images on arbitrary timescales (down to the 10\,s telescope integration time). These images are then formed into cubes and basic source statistics are calculated to enable a search for transients. This pipeline has historically been used on archival data, but has recently been deployed as part of the real-time telescope operations (Wang et al. in prep.).
    \item TRON - Transient Radio Observations for Newbies is a fast imaging pipeline developed in the context of mining the MeerKAT archive for transients and variable sources \citep{smirnov2025mining}. TRON creates a model-subtracted snapshot cube at raw time resolution, using either {\sc WSClean} \citep{wsclean} or the new {\sc pfb-imaging} package \citep{pfb}, stored in chunked format using {\sc XArray/Zarr}\footnote{\url{https://xarray.dev}} for efficient access. The data can be imaged over a number of subbands (which optimizes sensitivity to chromatic sources such as scintillators); this version of the pipeline is called SpecTRON. It then applies a peak-finding heuristic called {\sc Breifast} to search for candidate transients at the raw timescale, and/or in forward-difference snapshots and/or at longer timescales (by applying a convolutional filter in time) and/or in variance maps, with a number of heuristics to filter out false-positives. A catalogue of lightcurves for all detections (and all deep continuum sources) is automatically produced, as well as an optional set of dynamic spectra generated using the {\sc RIMS} package\footnote{\url{https://github.com/saopicc/RIMS}}.
    
    \item MWA fast imaging transient detection pipeline - Described by \cite{horvath2025long}, this pipeline is optimised for finding transients on timescales of a few seconds to a few minutes with the MWA. It produces a model-subtracted snapshot cube and applies three filters through the time axis, optimising for detection of delta-function like transients, high-variance pixels, and signatures resembling a pulse. These are uploaded to a classifier for human inspection.
\end{itemize}

\subsection{Filtering light curves and transient candidates}

Once light curves have been created and stored in the associated database, they are filtered using a range of strategies. New sources are considered transient candidates while persistent sources are checked for variability. 

\begin{figure*}
    \centering
    \includegraphics[width=0.9\textwidth]{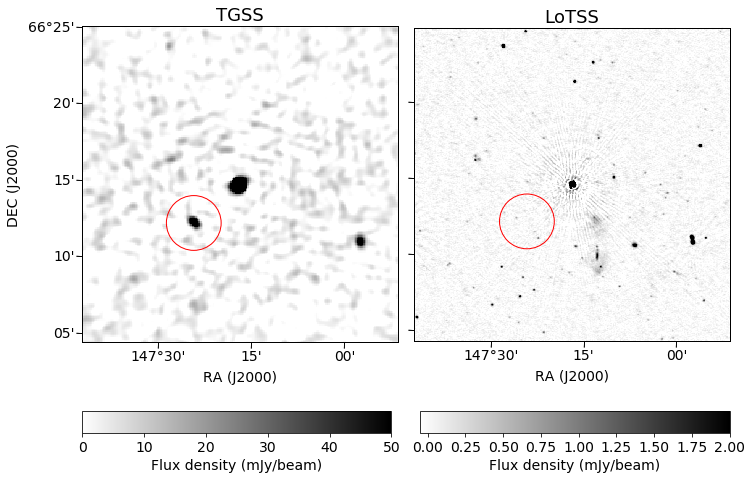}
    \protect \caption{
     TGSS ADR1 (left) and LoTSS DR2 (right) images of the AGN 4C66.09 shown in the centre of the images. The red circle shows the location of the sidelobe source (TGSSADR J094940.0+661228) that was initially interpreted as a transient candidate. Figure from \cite{de2021limits}.}
    \label{fig:sidelobe_example}
\end{figure*}

There are a number of artifacts that can lead to false positive transient detection or false positive variability in sources. Key sources of false positives include:
\begin{itemize}
    \item Sidelobes in radio images can appear as compact point sources in radio images, and can therefore unjustly be interpreted as transient candidates. These sidelobe sources are most prominent around bright sources, where a bright main central part of the point spread function (to model the bright source) also implies a relatively bright sidelobe of the point spread function. Figure \ref{fig:sidelobe_example} shows an example of a sidelobe highlighted by the red circle in the TGSS survey \citep{intema2017TGSSADR} around the 4.3 Jy AGN 4C66.09. Upon further investigation and comparison to the LoTSS DR2 survey \citep{shimwell2022lofar}, this source is determined to be spurious. The LoTSS image also shows residual structure around the central AGN. This structure encapsulates the shape of the PSF, determined by the UV-coverage of the radio interferometer. The amount of structure around bright sources is furthermore determined by the skymodel that is used for calibration. For example, if a point source model is adopted for a slightly resolved source, this can result in features around the slightly resolved source. This can be remedied by consecutive rounds of self-calibration \citep[see e.g.][]{stewart2016lofar}. 
    One strategy to reduce the impact of sidelobes on a transient search is to exclude a region around bright sources from the transient search. This can for example, be done by studying the noise distributions around bright sources and defining a radial filter \citep[see eg. Figure 8 in][]{de2021limits}. 
    \item Planes and satelites both emit and reflect radio emission that can cause false positive new source detections across the field as the source moves, particularly in fast imaging strategies. These false positives are likely to increase with the advent of constellation satellites. Simple strategies can be used to filter these out, such as searching for new source detections that are close together in both space and time \citep{kuiack2021}.
    \item Extended sources can also lead to false positive new source detections. This is due to source finders identifying different components of the extended emission from image to image, depending on e.g. correlated noise in the region. However, extended sources are unlikely to vary on the timescales that key transient surveys are conducted. Therefore, excluding known extended sources from transient analysis can prevent these false positives. The key caveat is that a transient source occurring on an extended galaxy could be rejected by this method. Alternatively, shorter baselines could be excluded when making images, giving lower resolution and thus reduce the amount of extended sources in the images.
\end{itemize}

\begin{figure*}
    \centering
    \includegraphics[width=0.7\textwidth]{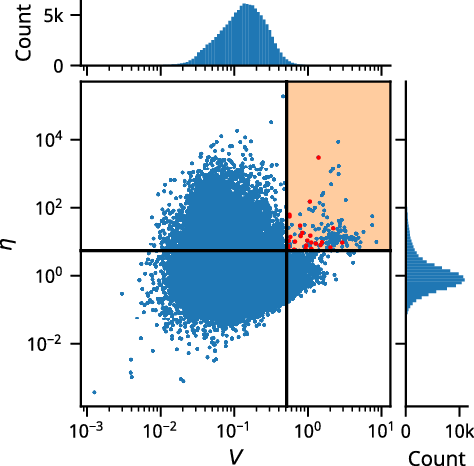}
    \protect \caption{This figure shows the two variability parameters, $\eta$ and $V$, determined from the light curves of all sources observed in a given dataset. The variability parameters each show an approximately Gaussian distribution (shown above and to the left of the plot). Sources that are deviant from both distributions by more than $2\sigma$ (shown by the solid black lines and orange shaded region) are candidate transient sources. The red data points are confirmed transient sources. Figure from \cite{murphy2021}}
    \label{fig:variabilityplot}
\end{figure*}

Often, the transient pipelines calculate key variability parameters such as the reduced weighted $\chi ^2$ fit of the light curve to a straight line ($\eta$) and the coefficient of variation giving the magnitude of the variability \citep[$V$, see e.g. equations 35 and 36 in ][]{swinbank2015}. Making the assumption that the majority of the sources are not variable, variable sources can be identified as being outliers to the distribution. In Figure \ref{fig:variabilityplot}, we show an example dataset where variable sources are defined as being $>2\sigma$ outliers from the full population. We note that there are caveats with this method as some variables are being missed due to various systematic effects in calculating the variability parameters \citep[e.g. see discussion in ][]{Rowlinson2019}. These systematic effects can be modeled and accounted for by adapting this analysis technique (see Valdata et al. in prep). Additionally, Gaussian processes can be used to analyse light curves that have challenges such as sparse and irregular sampling \citep{fu2025}.

Storing light curves, instead of classifying as part of the transient pipeline, enables astronomers to develop and apply new advanced analysis techniques on these catalogues when the images may no longer be available. For instance, in Section 3.4, we consider the application of citizen science and machine learning techniques to identify interesting source that are missed by the simple statistical techniques using the variability parameters.

\subsection{Special considerations for subtraction images}

Filtering applies to both short snapshot (section \ref{sec:short_snapshot_surveys}) and long surveys (section \ref{sec:long_surveys}), but extra care is needed for sidelobes around bright sources in short snapshots. This is because (1) sky-models, derived from long observations with better uv-coverage, do not remove sidelobes effectively in short integrations, leaving artifacts around bright sources; and (2) short snapshots are usually not cleaned, so unsubtracted sources appear as dirty beams with full sidelobes.
The excellent instantaneous $u,v$-coverage of the SKAO telescope will mitigate some of these difficulties. Additionally, the use of dishes also leads to cleaner subtractions \citep[see e.g.][]{fijma2024new}.

\subsection{Role of citizen science and machine learning}\label{sec:classification}

The scale of data expected from the SKAO necessitates fast, flexible methods to address the challenge of finding and characterising transients. The advent and maturity of machine learning and other big data techniques therefore provide a useful toolkit to aid in removing artifacts, classifying transients and optimising our pipelines, as is now standard practice for analogous optical sky surveys \citep[e.g.][]{Killestein2021_GOTO_R/B, Weston2024_ATLAS_R/B}. 
For example, \cite{Rowlinson2019} applied machine learning methods to simulated and observed LOFAR data in order to explore parameter tuning for particular survey strategies, which will be invaluable when balancing false positive rates against completeness with the sensitive observations of the SKAO. More recently \cite{Andersson2025} made use of anomaly detection algorithms - a kind of unsupervised machine learning - to preferentially select verified transients and reduce the required search volume by over an order of magnitude.

Citizen science is another strategy employable when faced with such large data volumes, with the additional benefits associated with engagement, outreach and learning opportunities. \cite{Andersson2023_citsci} demonstrate how volunteers find transients and variables from the ThunderKAT survey without the need for any choice in variability parameters or thresholding. Furthermore, this labelled dataset has proved useful as a testbed against which to compare other strategies. For example, \cite{fu2025} use the same citizen science data to explore the application of Gaussian processes to radio transients, whilst \cite{Andersson2025} use the volunteer votes to construct a ground truth against which to compare anomaly detection algorithms. However, at the SKAO AA* sensitivity predictions, the rate of variable and transient sources will likely be too great for such citizen science efforts to be employed on the entire SKAO datastream. Nevertheless, having a small, labeled dataset allows for a way to test novel techniques. 

Armed with these techniques that have now been benchmarked with and applied to data from SKA precursors and pathfinders, we are now moving into a regime where finding transients of interest will become overwhelming. The final piece of this puzzle is to integrate these techniques into  a real-time, low-latency system so that characterisation can be performed whilst transients are still active.

\section{Triggering}

The SKAO telescopes will produce a range of transient candidate type science data products through the real-time pulsar sub-systems and also from the fast imaging pipeline. These science data products can be used to generate alerts to trigger the SKAO telescopes for advanced analysis, new ToO observations or observations by external facilities. SKA-Mid and SKA-Low will have limited ToO availability from Cycle 0.

The SKAO telescopes are also able to save voltage data from the transient buffer upon triggering as part of the transient search pipelines that would provide high-time resolution visibilities for localisation and dynamic spectrum analysis of the transient candidate.

\subsection{Advanced Analysis}

Following the detection of new candidate transient and variable sources, a number of advanced analysis steps are used to confirm that a detection is real and to determine its properties. In the SKAO era, where the data rates prohibit long term storage of visibilities, these analysis steps will need to be conducted before the visibilities are deleted. 

By self triggering, ideally in real time, transient buffer data around key transient detections could be saved to disk enabling the advanced analysis techniques. Additionally, by identifying transients in real-time, self-triggered SKAO observations could be started to follow-up these events. Here we outline a range of typical advanced analysis steps that are used to analyse transient candidates.

\subsubsection{Re-imaging}

Following the detection of a new transient source, particularly from fast snapshot subtraction imaging surveys, it is essential to conduct re-imaging to confirm the transient detection. 

The key purpose is to confirm that transient detections are reliable and is ideally conducted before a transient alert is distributed. For subtraction images, re-imaging is required to confirm the source is real and to accurately determine its flux density. Often, transient candidates are re-imaged with different imaging strategies to confirm the source is not an artefact of given imaging settings. Additionally, re-imaging with different array configurations (e.g. baseline length constraints, frequency bands or polarisation) can be used to reject artefacts from sidelobes.

The second purpose is to measure useful properties of the source that can be used to constrain emission mechanisms and source type. Often sources are re-imaged with higher time resolution to search for shorter duration variability, higher spectral resolution to determine the spectrum and full Stokes imaging to identify any polarised emission.

\subsubsection{Dynamic Spectra}

At the time of writing, image-plane transient detection is typically performed on as coarse a timescale as is still sensitive to the transients of interest, in order to maximise signal-to-noise, and minimise computational cost. It is also often performed while averaging over all frequency channels without dedispersion, since many sources detected in the image plane to date have pulses long enough that dispersion does not appreciably reduce the signal-to-noise of the frequency-averaged light curve, and again, this minimises computational cost (although, see section~\ref{sec:dedm}). We note that this method can be biased against highly dispersed, short duration transients. Stokes I and Stokes V are common search dimensions, since Q and U are Faraday-rotated by intervening $B_\parallel$ fields and would necessitate costly per-channel images.

After a detection has been made, however, the richest information comes from exploring the data at full resolution across time, frequency, and polarisation. An increasingly popular method involves phase-rotating the visibilities to the detected source, and then averaging over all baselines. This is essentially the same beamforming operation as performed by traditional time domain astronomers, effectively placing a dirty, naturally-weighted synthesised beam at the location of the source. If the transient was never cleaned in the detection step, or if its model components are restored, then this method returns a dynamic spectrum of the source in full spectro-temporal-polarimetric detail (if the source has been cleaned, some level of mean subtraction will occur). A primary beam correction can also be applied at this stage, in order to convert from instrumental to celestial Stokes parameters. This method is becoming increasingly popular and there are at least two extant software implementations: \textsc{DSTools}\footnote{\url{https://github.com/askap-vast/dstools}} and \textsc{RIMS}\footnote{\url{https://github.com/saopicc/RIMS}}

In the SKAO era, where the visibilities will not be archived long-term, it will be vitally important to detect transients in real-time, then form and store these dynamic spectra for future inspection and classification. However, these are large data products compared to light curves and will need to be classified promptly to avoid an overwhelming data management problem (see section~\ref{sec:classification}). This is similar to the near-real time classification problem needed to keep pace with pulsar surveys with SKAO precursors \citep[see e.g.][]{2023MNRAS.524.1291P}.

\subsubsection{Image-plane dedispersion}\label{sec:dedm}
At low frequencies (150 MHz), dispersion is a dominant effect especially for transient sources. Even for moderate values of the dispersion measure (that measures the amount of smear of an impulse across the band) can lead to a reduction in the sensitivity of image plane searches, even for long-period transients. At LOFAR frequencies of 150~MHz (SKA-LOW regime), dispersion smear can be as large as 10 seconds for DMs larger than 50~pc~cm$^{-3}$. This renders an image-plane search insensitive to sources such as Long-Period Transients in the Galactic Plane as the pulse widths are can be shorter than the dispersion smearing at these frequencies. Hence, it is important to devise a dedispersion search for transients for SKA-low stations. The images are first corrected for dispersive effects across frequency channels before adding them and running a search thus, allowing us to look any highly dispersed transients. This image-plane dedispersion technique can be run on detected transients to determine if they are dispersed and what their dispersion measure is. This methodology has already been put to good use in searches for long period transients using LOFAR \citep[e.g.][]{deruiter2025}.

In the SKAO era, it would be important to consider two things: 1) the time resolution and the frequency resolution of the output images as the frequency resolution will dictate the upper limit on the dispersion measure the search would be sensitive to and 2) time requirement for the search. Currently, a heterogeneous pipeline has been deployed for the image plane searches with LOFAR that uses GPUs and CPUs to dedisperse the images giving a significant boost to the performance. Similar techniques can be implemented for the searches using the SKAO telescopes as a triggered analysis on detected transients and as a standard search strategy given sufficient compute.

\subsection{Alerts to the community}\label{sec:alerts}

Upon the detection (and possibly vetting) of a transient candidate, an alert will be rapidly disseminated to the global astronomical community to facilitate follow-up observations and studies. To ensure interoperability and seamless integration with existing systems, these alerts will be issued in the VOEvent format~\citep{2011ivoa.spec.0711S}, an established standard from the International Virtual Observatory Alliance (IVOA). The distribution of these alerts will likely be handled by the SKAO operating its own distribution broker to ensure reliable and low-latency delivery.

This infrastructure will allow the SKAO transient alert stream to be readily incorporated into the burgeoning global ecosystem of time-domain and multi-messenger (TDAMM) astronomy platforms. This includes major alert brokers developed for facilities like the Vera C. Rubin Observatory, enabling immediate cross-correlation of SKAO radio detections with optical alerts from the Legacy Survey of Space and Time \citep[LSST; ][]{lsst_brokers} and data from other multi-wavelength and multi-messenger observatories. Such near-real-time data fusion significantly enhances the scientific value of SKAO alerts, providing crucial context for classifying events and triggering rapid follow-up observations. Furthermore, the SKAO alert stream will be integrated into platforms like Astro-COLIBRI~\citep{2021ApJS..256....5R}, which provides a comprehensive, high-level overview of all ongoing transient activity across the electromagnetic spectrum and beyond.

Finally, while the initial VOEvents will contain essential information about the detection (e.g., position, flux density, time), dedicated SKAO-based information and data platforms will need to be made available via the SRCNet. These will provide the community with access to more detailed information for each event. This should include the source-subtracted snapshot image, the dynamic spectrum of the event, the visibility data around the time of the transient or the snapshot images around the transient event including all the sources and associated metadata.

\section{Conclusions}

Working with the SKA pathfinder facilities and collaborating with teams from other science cases over the past decade, the radio transient community has demonstrated the high quality results obtainable from commensal transient searches in the image plane. As shown by the detection of Long Period Transients \citep{Qiu01.2026.SKA} in commensal data, these searches are vital for detecting rare transients and the unknown \citep{AlexAndersson01.2026.SKA}. 

This chapter draws together the state of the art techniques for conducting transient and variability searches from durations of seconds to years. Imaging strategies, especially for short duration snapshots, have been optimised for speed and data quality. Transient detection pipelines have been developed that can build light curves of sources detected in the fields. These light curves can be processed in a range of different ways, including machine learning techniques, to pull out the interesting sources. Finally, we have identified key follow-up strategies to reprocess the observational radio data to pull out key source features and determined how to share the detection of these exciting sources with the multi-messenger transient community to enable timely follow-up observations.

\section*{Acknowledgments}
DD and IdR are supported by the Australian Research Council Centre of Excellence for Gravitational Wave Discovery (OzGrav), project number CE230100016.

KR acknowleges funding from a UKRI-STFC grant (SKA-NIPS, no. ST/Z510439/1).

AR acknowledges funding from the European Research Council (ERC) under the European Union’s Horizon research and innovation programme (‘QuickBlitz’; grant agreement number 101170284), funding from the NWO Aspasia grant (number 015.016.033), and support through the project CORTEX (project number NWA.1160.18.316) of the research programme NWA-ORC, which is (partly) financed by the Dutch Research Council (NWO). Views and opinions expressed are however those of the author(s) only and do not necessarily reflect those of the European Union or the European Research Council. Neither the European Union nor the granting authority can be held responsible for them.

FS acknowledges ANR (French National Research Agency) for its support of the project "Multi-messenger observations of the Transient Sky (MOTS)" under grant no. ANR-22-CE31-0012.

OMS’s research is supported by the South African Research Chairs Initiative of the Department of Science, Technology and Innovation and the National Research Foundation (grant No. 81737).

\bibliographystyle{abbrvnat-maxbibnames4}
\bibliography{chapter} 

@ARTICLE{deruiter2025,
       author = {{de Ruiter}, I. and {Rajwade}, K.~M. and {Bassa}, C.~G. and {Rowlinson}, A. and {Wijers}, R.~A.~M.~J. and {Kilpatrick}, C.~D. and {Stefansson}, G. and {Callingham}, J.~R. and {Hessels}, J.~W.~T. and {Clarke}, T.~E. and {Peters}, W. and {Wijnands}, R.~A.~D. and {Shimwell}, T.~W. and {ter Veen}, S. and {Morello}, V. and {Zeimann}, G.~R. and {Mahadevan}, S.},
        title = "{Sporadic radio pulses from a white dwarf binary at the orbital period}",
      journal = {Nature Astronomy},
         year = 2025,
        month = may,
       volume = {9},
        pages = {672-684},
          doi = {10.1038/s41550-025-02491-0},
archivePrefix = {arXiv},
       eprint = {2408.11536},
 primaryClass = {astro-ph.HE},
       adsurl = {https://ui.adsabs.harvard.edu/abs/2025NatAs...9..672D}
}

@incollection{AlexAndersson01.2026.SKA, author = {Alex Andersson and author2 and author3 and author4 and author5},title = {},year = {2026},publisher = {},note = {arXiv search: Report number AASKAII/AlexAndersson01},booktitle = {Advancing Astrophysics with the SKA -- II (AASKAII)}}

@incollection{Qiu01.2026.SKA, author = {Hao Qiu and author2 and author3 and author4 and author5},title = {},year = {2026},publisher = {},note = {arXiv search: Report number AASKAII/Qiu01},booktitle = {Advancing Astrophysics with the SKA -- II (AASKAII)}}

@incollection{GemmaAnderson01.2026.SKA, author = {Gemma E. Anderson and author2 and author3 and author4 and author5},title = {},year = {2026},publisher = {},note = {arXiv search: Report number AASKAII/GemmaAnderson01},booktitle = {Advancing Astrophysics with the SKA -- II (AASKAII)}}

@incollection{Colombo01.2026.SKA, author = {Alberto Colombo and author2 and author3 and author4 and author5},title = {},year = {2026},publisher = {},note = {arXiv search: Report number AASKAII/Colombo01},booktitle = {Advancing Astrophysics with the SKA -- II (AASKAII)}}

@incollection{Beri01.2026.SKA, author = {Aru Beri and author2 and author3 and author4 and author5},title = {},year = {2026},publisher = {},note = {arXiv search: Report number AASKAII/Beri01},booktitle = {Advancing Astrophysics with the SKA -- II (AASKAII)}}

@incollection{Driessen01.2026.SKA, author = {Laura N. Driessen and author2 and author3 and author4 and author5},title = {},year = {2026},publisher = {},note = {arXiv search: Report number AASKAII/Driessen01},booktitle = {Advancing Astrophysics with the SKA -- II (AASKAII)}}

@incollection{Curtin01.2026.SKA, author = {Alice P. Curtin and author2 and author3 and author4 and author5},title = {},year = {2026},publisher = {},note = {arXiv search: Report number AASKAII/Curtin01},booktitle = {Advancing Astrophysics with the SKA -- II (AASKAII)}}

@incollection{Caleb02.2026.SKA, author = {Manisha Caleb and author2 and author3 and author4 and author5},title = {},year = {2026},publisher = {},note = {arXiv search: Report number AASKAII/Caleb02},booktitle = {Advancing Astrophysics with the SKA -- II (AASKAII)}}

@ARTICLE{wang2021,
       author = {{Wang}, Yuanming and {Tuntsov}, Artem and {Murphy}, Tara and {Lenc}, Emil and {Walker}, Mark and {Bannister}, Keith and {Kaplan}, David L. and {Mahony}, Elizabeth K.},
        title = "{ASKAP observations of multiple rapid scintillators reveal a degrees-long plasma filament}",
      journal = {\mnras},
         year = 2021,
        month = apr,
       volume = {502},
       number = {3},
        pages = {3294-3311},
          doi = {10.1093/mnras/stab139},
archivePrefix = {arXiv},
       eprint = {2101.06048},
 primaryClass = {astro-ph.GA},
       adsurl = {https://ui.adsabs.harvard.edu/abs/2021MNRAS.502.3294W}
}

@ARTICLE{chastain2022,
       author = {{Chastain}, S.~I. and {van der Horst}, A.~J. and {Carbone}, D.},
        title = "{Transient simulations for radio surveys}",
      journal = {Astronomy and Computing},
         year = 2022,
        month = jul,
       volume = {40},
          eid = {100629},
        pages = {100629},
          doi = {10.1016/j.ascom.2022.100629},
archivePrefix = {arXiv},
       eprint = {2208.00965},
 primaryClass = {astro-ph.HE},
       adsurl = {https://ui.adsabs.harvard.edu/abs/2022A&C....4000629C}
}

@ARTICLE{fu2025,
       author = {{Fu}, Shih Ching and {Bahramian}, Arash and {Phatak}, Aloke and {Miller-Jones}, James C.~A. and {Rakshit}, Suman and {Andersson}, Alexander and {Fender}, Robert and {Woudt}, Patrick A.},
        title = "{New Metrics for Identifying Variables and Transients in Large Astronomical Surveys}",
      journal = {\apj},
         year = 2025,
        month = oct,
       volume = {992},
       number = {1},
          eid = {109},
        pages = {109},
          doi = {10.3847/1538-4357/adfb79},
archivePrefix = {arXiv},
       eprint = {2508.09441},
 primaryClass = {astro-ph.HE},
       adsurl = {https://ui.adsabs.harvard.edu/abs/2025ApJ...992..109F}
}

@ARTICLE{kuiack2021,
       author = {{Kuiack}, Mark and {Wijers}, Ralph A.~M.~J. and {Shulevski}, Aleksandar and {Rowlinson}, Antonia and {Huizinga}, Folkert and {Molenaar}, Gijs and {Prasad}, Peeyush},
        title = "{The AARTFAAC 60 MHz transients survey}",
      journal = {\mnras},
         year = 2021,
        month = aug,
       volume = {505},
       number = {2},
        pages = {2966-2974},
          doi = {10.1093/mnras/stab1504},
archivePrefix = {arXiv},
       eprint = {2003.13289},
 primaryClass = {astro-ph.HE},
       adsurl = {https://ui.adsabs.harvard.edu/abs/2021MNRAS.505.2966K}
}

@ARTICLE{carbone2018,
       author = {{Carbone}, D. and {Garsden}, H. and {Spreeuw}, H. and {Swinbank}, J.~D. and {van der Horst}, A.~J. and {Rowlinson}, A. and {Broderick}, J.~W. and {Rol}, E. and {Law}, C. and {Molenaar}, G. and {Wijers}, R.~A.~M.~J.},
        title = "{PySE: Software for extracting sources from radio images}",
      journal = {Astronomy and Computing},
         year = 2018,
        month = apr,
       volume = {23},
          eid = {92},
        pages = {92},
          doi = {10.1016/j.ascom.2018.02.003},
archivePrefix = {arXiv},
       eprint = {1802.09604},
 primaryClass = {astro-ph.IM},
       adsurl = {https://ui.adsabs.harvard.edu/abs/2018A&C....23...92C}
}

@ARTICLE{rowlinson2016,
       author = {{Rowlinson}, A. and {Bell}, M.~E. and {Murphy}, T. and {Trott}, C.~M. and {Hurley-Walker}, N. and {Johnston}, S. and {Tingay}, S.~J. and {Kaplan}, D.~L. and {Carbone}, D. and {Hancock}, P.~J. and {Feng}, L. and {Offringa}, A.~R. and {Bernardi}, G. and {Bowman}, J.~D. and {Briggs}, F. and {Cappallo}, R.~J. and {Deshpande}, A.~A. and {Gaensler}, B.~M. and {Greenhill}, L.~J. and {Hazelton}, B.~J. and {Johnston-Hollitt}, M. and {Lonsdale}, C.~J. and {McWhirter}, S.~R. and {Mitchell}, D.~A. and {Morales}, M.~F. and {Morgan}, E. and {Oberoi}, D. and {Ord}, S.~M. and {Prabu}, T. and {Udaya Shankar}, N. and {Srivani}, K.~S. and {Subrahmanyan}, R. and {Wayth}, R.~B. and {Webster}, R.~L. and {Williams}, A. and {Williams}, C.~L.},
        title = "{Limits on Fast Radio Bursts and other transient sources at 182 MHz using the Murchison Widefield Array}",
      journal = {\mnras},
         year = 2016,
        month = jun,
       volume = {458},
       number = {4},
        pages = {3506-3522},
          doi = {10.1093/mnras/stw451},
archivePrefix = {arXiv},
       eprint = {1602.07544},
 primaryClass = {astro-ph.HE},
       adsurl = {https://ui.adsabs.harvard.edu/abs/2016MNRAS.458.3506R}
}

@ARTICLE{murphy2021,
       author = {{Murphy}, Tara and {Kaplan}, David L. and {Stewart}, Adam J. and {O'Brien}, Andrew and {Lenc}, Emil and {Pintaldi}, Sergio and {Pritchard}, Joshua and {Dobie}, Dougal and {Fox}, Archibald and {Leung}, James K. and {An}, Tao and {Bell}, Martin E. and {Broderick}, Jess W. and {Chatterjee}, Shami and {Dai}, Shi and {d'Antonio}, Daniele and {Doyle}, Gerry and {Gaensler}, B.~M. and {Heald}, George and {Horesh}, Assaf and {Jones}, Megan L. and {McConnell}, David and {Moss}, Vanessa A. and {Raja}, Wasim and {Ramsay}, Gavin and {Ryder}, Stuart and {Sadler}, Elaine M. and {Sivakoff}, Gregory R. and {Wang}, Yuanming and {Wang}, Ziteng and {Wheatland}, Michael S. and {Whiting}, Matthew and {Allison}, James R. and {Anderson}, C.~S. and {Ball}, Lewis and {Bannister}, K. and {Bock}, D.~C. -J. and {Bolton}, R. and {Bunton}, J.~D. and {Chekkala}, R. and {Chippendale}, A.~P. and {Cooray}, F.~R. and {Gupta}, N. and {Hayman}, D.~B. and {Jeganathan}, K. and {Koribalski}, B. and {Lee-Waddell}, K. and {Mahony}, Elizabeth K. and {Marvil}, J. and {McClure-Griffiths}, N.~M. and {Mirtschin}, P. and {Ng}, A. and {Pearce}, S. and {Phillips}, C. and {Voronkov}, M.~A.},
        title = "{The ASKAP Variables and Slow Transients (VAST) Pilot Survey}",
      journal = {\pasa},
         year = 2021,
        month = oct,
       volume = {38},
          eid = {e054},
        pages = {e054},
          doi = {10.1017/pasa.2021.44},
archivePrefix = {arXiv},
       eprint = {2108.06039},
 primaryClass = {astro-ph.HE},
       adsurl = {https://ui.adsabs.harvard.edu/abs/2021PASA...38...54M}
}

@ARTICLE{deruiter2024,
       author = {{de Ruiter}, Iris and {Meyers}, Zachary S. and {Rowlinson}, Antonia and {Shimwell}, Timothy W. and {Ruhe}, David and {Wijers}, Ralph A.~M.~J.},
        title = "{Transient study using LoTSS - framework development and preliminary results}",
      journal = {\mnras},
         year = 2024,
        month = jul,
       volume = {531},
       number = {4},
        pages = {4805-4822},
          doi = {10.1093/mnras/stae1458},
archivePrefix = {arXiv},
       eprint = {2311.07394},
 primaryClass = {astro-ph.HE},
       adsurl = {https://ui.adsabs.harvard.edu/abs/2024MNRAS.531.4805D}
}

@ARTICLE{swinbank2015,
       author = {{Swinbank}, John D. and {Staley}, Tim D. and {Molenaar}, Gijs J. and {Rol}, Evert and {Rowlinson}, Antonia and {Scheers}, Bart and {Spreeuw}, Hanno and {Bell}, Martin E. and {Broderick}, Jess W. and {Carbone}, Dario and {Garsden}, Hugh and {van der Horst}, Alexander J. and {Law}, Casey J. and {Wise}, Michael and {Breton}, Rene P. and {Cendes}, Yvette and {Corbel}, St{\'e}phane and {Eisl{\"o}ffel}, Jochen and {Falcke}, Heino and {Fender}, Rob and {Grie{\ss}meier}, Jean-Mathias and {Hessels}, Jason W.~T. and {Stappers}, Benjamin W. and {Stewart}, Adam J. and {Wijers}, Ralph A.~M.~J. and {Wijnands}, Rudy and {Zarka}, Philippe},
        title = "{The LOFAR Transients Pipeline}",
      journal = {Astronomy and Computing},
         year = 2015,
        month = jun,
       volume = {11},
        pages = {25-48},
          doi = {10.1016/j.ascom.2015.03.002},
archivePrefix = {arXiv},
       eprint = {1503.01526},
 primaryClass = {astro-ph.IM},
       adsurl = {https://ui.adsabs.harvard.edu/abs/2015A&C....11...25S}
}

@ARTICLE{hurleywalker2022,
       author = {{Hurley-Walker}, N. and {Zhang}, X. and {Bahramian}, A. and {McSweeney}, S.~J. and {O'Doherty}, T.~N. and {Hancock}, P.~J. and {Morgan}, J.~S. and {Anderson}, G.~E. and {Heald}, G.~H. and {Galvin}, T.~J.},
        title = "{A radio transient with unusually slow periodic emission}",
      journal = {\nat},
         year = 2022,
        month = jan,
       volume = {601},
       number = {7894},
        pages = {526-530},
          doi = {10.1038/s41586-021-04272-x},
archivePrefix = {arXiv},
       eprint = {2503.08033},
 primaryClass = {astro-ph.HE},
       adsurl = {https://ui.adsabs.harvard.edu/abs/2022Natur.601..526H}
}

@ARTICLE{metzger2015,
       author = {{Metzger}, Brian D. and {Williams}, P.~K.~G. and {Berger}, Edo},
        title = "{Extragalactic Synchrotron Transients in the Era of Wide-field Radio Surveys. I. Detection Rates and Light Curve Characteristics}",
      journal = {\apj},
         year = 2015,
        month = jun,
       volume = {806},
       number = {2},
          eid = {224},
        pages = {224},
          doi = {10.1088/0004-637X/806/2/224},
archivePrefix = {arXiv},
       eprint = {1502.01350},
 primaryClass = {astro-ph.HE},
       adsurl = {https://ui.adsabs.harvard.edu/abs/2015ApJ...806..224M}
}

@ARTICLE{lorimer2007,
       author = {{Lorimer}, D.~R. and {Bailes}, M. and {McLaughlin}, M.~A. and {Narkevic}, D.~J. and {Crawford}, F.},
        title = "{A Bright Millisecond Radio Burst of Extragalactic Origin}",
      journal = {Science},
         year = 2007,
        month = nov,
       volume = {318},
       number = {5851},
        pages = {777},
          doi = {10.1126/science.1147532},
archivePrefix = {arXiv},
       eprint = {0709.4301},
 primaryClass = {astro-ph},
       adsurl = {https://ui.adsabs.harvard.edu/abs/2007Sci...318..777L}
}

@ARTICLE{prentice2018,
       author = {{Prentice}, S.~J. and {Maguire}, K. and {Smartt}, S.~J. and {Magee}, M.~R. and {Schady}, P. and {Sim}, S. and {Chen}, T. -W. and {Clark}, P. and {Colin}, C. and {Fulton}, M. and {McBrien}, O. and {O'Neill}, D. and {Smith}, K.~W. and {Ashall}, C. and {Chambers}, K.~C. and {Denneau}, L. and {Flewelling}, H.~A. and {Heinze}, A. and {Holoien}, T.~W. -S. and {Huber}, M.~E. and {Kochanek}, C.~S. and {Mazzali}, P.~A. and {Prieto}, J.~L. and {Rest}, A. and {Shappee}, B.~J. and {Stalder}, B. and {Stanek}, K.~Z. and {Stritzinger}, M.~D. and {Thompson}, T.~A. and {Tonry}, J.~L.},
        title = "{The Cow: Discovery of a Luminous, Hot, and Rapidly Evolving Transient}",
      journal = {\apjl},
         year = 2018,
        month = sep,
       volume = {865},
       number = {1},
          eid = {L3},
        pages = {L3},
          doi = {10.3847/2041-8213/aadd90},
archivePrefix = {arXiv},
       eprint = {1807.05965},
 primaryClass = {astro-ph.HE},
       adsurl = {https://ui.adsabs.harvard.edu/abs/2018ApJ...865L...3P}
}

@ARTICLE{klebesadel1973,
       author = {{Klebesadel}, Ray W. and {Strong}, Ian B. and {Olson}, Roy A.},
        title = "{Observations of Gamma-Ray Bursts of Cosmic Origin}",
      journal = {\apjl},
         year = 1973,
        month = jun,
       volume = {182},
        pages = {L85},
          doi = {10.1086/181225},
       adsurl = {https://ui.adsabs.harvard.edu/abs/1973ApJ...182L..85K}
}

@ARTICLE{Andersson2025,
       author = {{Andersson}, Alex and {Lintott}, Chris and {Fender}, Rob and {Lochner}, Michelle and {Woudt}, Patrick and {van den Eijnden}, Jakob and {van der Horst}, Alexander and {Horesh}, Assaf and {Saikia}, Payaswini and {Sivakoff}, Gregory R. and {Tremou}, Lilia and {Vaccari}, Mattia},
        title = "{Finding radio transients with anomaly detection and active learning based on volunteer classifications}",
      journal = {\mnras},
         year = 2025,
        month = apr,
       volume = {538},
       number = {3},
        pages = {1397-1414},
          doi = {10.1093/mnras/staf336},
archivePrefix = {arXiv},
       eprint = {2410.01034},
 primaryClass = {astro-ph.IM},
       adsurl = {https://ui.adsabs.harvard.edu/abs/2025MNRAS.538.1397A}
}

@ARTICLE{Andersson2023_citsci,
       author = {{Andersson}, Alex and {Lintott}, Chris and {Fender}, Rob and {Bright}, Joe and {Carotenuto}, Francesco and {Driessen}, Laura and {Espinasse}, Mathilde and {Gasealahwe}, Kelebogile and {Heywood}, Ian and {van der Horst}, Alexander J. and {Motta}, Sara and {Rhodes}, Lauren and {Tremou}, Evangelia and {Williams}, David R.~A. and {Woudt}, Patrick and {Zhang}, Xian and {Bloemen}, Steven and {Groot}, Paul and {Vreeswijk}, Paul and {Giarratana}, Stefano and {Saikia}, Payaswini and {Andersson}, Jonas and {Ruiz Arroyo}, Lizzeth and {Baert}, Lo{\"\i}c and {Baumann}, Matthew and {Domainko}, Wilfried and {Eschweiler}, Thorsten and {Forsythe}, Tim and {Gaudenzi}, Sauro and {Ann Grenier}, Rachel and {Iannone}, Davide and {Lahoz}, Karla and {Melville}, Kyle J. and {De Sousa Nascimento}, Marianne and {Navarro}, Leticia and {Parthasarathi}, Sai and {Piilonen} and {Rahman}, Najma and {Smith}, Jeffrey and {Stewart}, B. and {Temoke}, Newton and {Tworek}, Chloe and {Whittle}, Isabelle},
        title = "{Bursts from Space: MeerKAT - the first citizen science project dedicated to commensal radio transients}",
      journal = {\mnras},
         year = 2023,
        month = aug,
       volume = {523},
       number = {2},
        pages = {2219-2235},
          doi = {10.1093/mnras/stad1298},
archivePrefix = {arXiv},
       eprint = {2304.14157},
 primaryClass = {astro-ph.HE},
       adsurl = {https://ui.adsabs.harvard.edu/abs/2023MNRAS.523.2219A}
}

@ARTICLE{Rowlinson2019,
       author = {{Rowlinson}, A. and {Stewart}, A.~J. and {Broderick}, J.~W. and {Swinbank}, J.~D. and {Wijers}, R.~A.~M.~J. and {Carbone}, D. and {Cendes}, Y. and {Fender}, R. and {van der Horst}, A. and {Molenaar}, G. and {Scheers}, B. and {Staley}, T. and {Farrell}, S. and {Grie{\ss}meier}, J. -M. and {Bell}, M. and {Eisl{\"o}ffel}, J. and {Law}, C.~J. and {van Leeuwen}, J. and {Zarka}, P.},
        title = "{Identifying transient and variable sources in radio images}",
      journal = {Astronomy and Computing},
         year = 2019,
        month = apr,
       volume = {27},
          eid = {111},
        pages = {111},
          doi = {10.1016/j.ascom.2019.03.003},
archivePrefix = {arXiv},
       eprint = {1808.07781},
 primaryClass = {astro-ph.IM},
       adsurl = {https://ui.adsabs.harvard.edu/abs/2019A&C....27..111R}
}

@ARTICLE{Killestein2021_GOTO_R/B,
       author = {{Killestein}, T.~L. and {Lyman}, J. and {Steeghs}, D. and {Ackley}, K. and {Dyer}, M.~J. and {Ulaczyk}, K. and {Cutter}, R. and {Mong}, Y. -L. and {Galloway}, D.~K. and {Dhillon}, V. and {O'Brien}, P. and {Ramsay}, G. and {Poshyachinda}, S. and {Kotak}, R. and {Breton}, R.~P. and {Nuttall}, L.~K. and {Pall{\'e}}, E. and {Pollacco}, D. and {Thrane}, E. and {Aukkaravittayapun}, S. and {Awiphan}, S. and {Burhanudin}, U. and {Chote}, P. and {Chrimes}, A. and {Daw}, E. and {Duffy}, C. and {Eyles-Ferris}, R. and {Gompertz}, B. and {Heikkil{\"a}}, T. and {Irawati}, P. and {Kennedy}, M.~R. and {Levan}, A. and {Littlefair}, S. and {Makrygianni}, L. and {Mata S{\'a}nchez}, D. and {Mattila}, S. and {Maund}, J. and {McCormac}, J. and {Mkrtichian}, D. and {Mullaney}, J. and {Rol}, E. and {Sawangwit}, U. and {Stanway}, E. and {Starling}, R. and {Str{\o}m}, P.~A. and {Tooke}, S. and {Wiersema}, K. and {Williams}, S.~C.},
        title = "{Transient-optimized real-bogus classification with Bayesian convolutional neural networks - sifting the GOTO candidate stream}",
      journal = {\mnras},
         year = 2021,
        month = may,
       volume = {503},
       number = {4},
        pages = {4838-4854},
          doi = {10.1093/mnras/stab633},
archivePrefix = {arXiv},
       eprint = {2102.09892},
 primaryClass = {astro-ph.IM},
       adsurl = {https://ui.adsabs.harvard.edu/abs/2021MNRAS.503.4838K}
}

@ARTICLE{Weston2024_ATLAS_R/B,
       author = {{Weston}, J.~G. and {Smith}, K.~W. and {Smartt}, S.~J. and {Tonry}, J.~L. and {Stevance}, H.~F.},
        title = "{Training a convolutional neural network for real-bogus classification in the ATLAS survey}",
      journal = {RAS Techniques and Instruments},
         year = 2024,
        month = jan,
       volume = {3},
       number = {1},
        pages = {385-399},
          doi = {10.1093/rasti/rzae027},
       adsurl = {https://ui.adsabs.harvard.edu/abs/2024RASTI...3..385W}
}

@article{caleb2022discovery,
  title={Discovery of a radio-emitting neutron star with an ultra-long spin period of 76 s},
  author={Caleb, Manisha and Heywood, Ian and Rajwade, Kaustubh and Malenta, Mateusz and Willem Stappers, Benjamin and Barr, Ewan and Chen, Weiwei and Morello, Vincent and Sanidas, Sotiris and Van Den Eijnden, Jakob and others},
  journal={Nature Astronomy},
  volume={6},
  number={7},
  pages={828--836},
  year={2022},
  publisher={Nature Publishing Group UK London}
}

@article{andrianjafy2023image,
  title={Image plane detection of FRB121102 with the MeerKAT radio telescope},
  author={Andrianjafy, JC and Heeralall-Issur, N and Deshpande, AA and Golap, Kumar and Woudt, P and Caleb, M and Barr, ED and Chen, W and Jankowski, F and Kramer, M and others},
  journal={Monthly Notices of the Royal Astronomical Society},
  volume={518},
  number={3},
  pages={3462--3474},
  year={2023},
  publisher={Oxford University Press}
}

@ARTICLE{murphy2025,
       author = {{Murphy}, Tara and {Kaplan}, David L.},
        title = "{The Dawes Review 13: A New Look at The Dynamic Radio Sky}",
      journal = {arXiv e-prints},
         year = 2025,
        month = nov,
          eid = {arXiv:2511.10785},
        pages = {arXiv:2511.10785},
          doi = {10.48550/arXiv.2511.10785},
archivePrefix = {arXiv},
       eprint = {2511.10785},
 primaryClass = {astro-ph.SR},
       adsurl = {https://ui.adsabs.harvard.edu/abs/2025arXiv251110785M}
}

@article{cordes2004dynamic,
  title={The dynamic radio sky},
  author={Cordes, James M and Lazio, T Joseph W and McLaughlin, MA},
  journal={New Astronomy Reviews},
  volume={48},
  number={11-12},
  pages={1459--1472},
  year={2004},
  publisher={Elsevier}
}

@article{nimmo2022burst,
  title={Burst timescales and luminosities as links between young pulsars and fast radio bursts},
  author={Nimmo, K and Hessels, JWT and Kirsten, F and Keimpema, A and Cordes, JM and Snelders, MP and Hewitt, DM and Karuppusamy, R and Archibald, AM and Bezrukovs, V and others},
  journal={Nature Astronomy},
  volume={6},
  number={3},
  pages={393--401},
  year={2022},
  publisher={Nature Publishing Group UK London}
}

@article{pietka2015variability,
  title={The variability time-scales and brightness temperatures of radio flares from stars to supermassive black holes},
  author={Pietka, M and Fender, RP and Keane, EF},
  journal={Monthly Notices of the Royal Astronomical Society},
  volume={446},
  number={4},
  pages={3687--3696},
  year={2015},
  publisher={Oxford University Press}
}

@article{wang2021askap,
  title={ASKAP observations of multiple rapid scintillators reveal a degrees-long plasma filament},
  author={Wang, Yuanming and Tuntsov, Artem and Murphy, Tara and Lenc, Emil and Walker, Mark and Bannister, Keith and Kaplan, David L and Mahony, Elizabeth K},
  journal={Monthly Notices of the Royal Astronomical Society},
  volume={502},
  number={3},
  pages={3294--3311},
  year={2021},
  publisher={Oxford University Press}
}

@article{liu2024testing,
  title={Testing the LSST Difference Image Analysis Pipeline Using Synthetic Source Injection Analysis},
  author={Liu, S and Wood-Vasey, WM and Armstrong, R and Narayan, G and S{\'a}nchez, BO and Dark Energy Science Collaboration and others},
  journal={The Astrophysical Journal},
  volume={967},
  number={1},
  pages={10},
  year={2024},
  publisher={IOP Publishing}
}

@article{smirnov2025mining,
  title={Mining the time axis with TRON--I. Millisecond pulsars in Omega Centauri, Terzan 5, and 47 Tucanae detected through MeerKAT interferometric imaging},
  author={Smirnov, Oleg M and Heywood, Ian and Geyer, Marisa and Myburgh, Talon and Tasse, Cyril and Kenyon, Jonathan S and Perkins, S J and Dawson, James and Bester, Hertzog L and Bright, Joe S and others},
  journal={Monthly Notices of the Royal Astronomical Society: Letters},
  volume={538},
  number={1},
  pages={L62--L68},
  year={2025},
  publisher={Oxford University Press}
}

@article{wang2023radio,
  title={Radio variable and transient sources on minute time-scales in the ASKAP pilot surveys},
  author={Wang, Yuanming and Murphy, Tara and Lenc, Emil and Mercorelli, Louis and Driessen, Laura and Pritchard, Joshua and Lao, Baoqiang and Kaplan, David L and An, Tao and Bannister, Keith W and others},
  journal={Monthly Notices of the Royal Astronomical Society},
  volume={523},
  number={4},
  pages={5661--5680},
  year={2023},
  publisher={Oxford University Press}
}

@article{de2024transient,
  title={Transient study using LoTSS--framework development and preliminary results},
  author={de Ruiter, Iris and Meyers, Zachary S and Rowlinson, Antonia and Shimwell, Timothy W and Ruhe, David and Wijers, Ralph AMJ},
  journal={Monthly Notices of the Royal Astronomical Society},
  volume={531},
  number={4},
  pages={4805--4822},
  year={2024},
  publisher={Oxford University Press}
}

@article{fijma2024new,
  title={A new method for short-duration transient detection in radio images: searching for transient sources in MeerKAT data of NGC 5068},
  author={Fijma, S and Rowlinson, A and Wijers, RAMJ and de Ruiter, I and de Blok, WJG and Chastain, S and van der Horst, AJ and Meyers, ZS and van der Meulen, K and Fender, R and others},
  journal={Monthly Notices of the Royal Astronomical Society},
  volume={528},
  number={4},
  pages={6985--6996},
  year={2024},
  publisher={Oxford University Press}
}

@ARTICLE{2021ApJS..256....5R,
       author = {{Reichherzer}, P. and {Sch{\"u}ssler}, F. and {Lefranc}, V. and {Yusafzai}, A. and {Alkan}, A.~K. and {Ashkar}, H. and {Becker Tjus}, J.},
        title = "{Astro-COLIBRI-The COincidence LIBrary for Real-time Inquiry for Multimessenger Astrophysics}",
      journal = {\apjs},
         year = 2021,
        month = sep,
       volume = {256},
       number = {1},
          eid = {5},
        pages = {5},
          doi = {10.3847/1538-4365/ac1517},
archivePrefix = {arXiv},
       eprint = {2109.01672},
 primaryClass = {astro-ph.IM},
       adsurl = {https://ui.adsabs.harvard.edu/abs/2021ApJS..256....5R}
}

@article{stewart2016lofar,
  title={LOFAR MSSS: detection of a low-frequency radio transient in 400 h of monitoring of the North Celestial Pole},
  author={Stewart, AJ and Fender, RP and Broderick, Jess W and Hassall, Tom E and Mu{\~n}oz-Darias, T and Rowlinson, Antonia and Swinbank, John D and Staley, Tim D and Molenaar, Gijs J and Scheers, Bart and others},
  journal={Monthly Notices of the Royal Astronomical Society},
  volume={456},
  number={3},
  pages={2321--2342},
  year={2016},
  publisher={The Royal Astronomical Society}
}

@article{de2021limits,
  title={Limits on long-time-scale radio transients at 150 MHz using the TGSS ADR1 and LoTSS DR2 catalogues},
  author={de Ruiter, Iris and Leseigneur, Guillaume and Rowlinson, Antonia and Wijers, Ralph AMJ and Drabent, Alexander and Intema, Huib T and Shimwell, Timothy W},
  journal={Monthly Notices of the Royal Astronomical Society},
  volume={508},
  number={2},
  pages={2412--2425},
  year={2021},
  publisher={Oxford University Press}
}

@article{shimwell2022lofar,
  title={The LOFAR two-metre sky survey-V. Second data release},
  author={Shimwell, TW and Hardcastle, MJ and Tasse, C and Best, PN and R{\"o}ttgering, HJA and Williams, WL and Botteon, A and Drabent, A and Mechev, A and Shulevski, A and others},
  journal={Astronomy \& astrophysics},
  volume={659},
  pages={A1},
  year={2022},
  publisher={EDP Sciences}
}

@ARTICLE{intema2017TGSSADR,
       author = {{Intema}, H.~T. and {Jagannathan}, P. and {Mooley}, K.~P. and
         {Frail}, D.~A.},
        title = "{The GMRT 150 MHz all-sky radio survey. First alternative data release TGSS ADR1}",
      journal = {\aap},
         year = "2017",
        month = "Feb",
       volume = {598},
          eid = {A78},
        pages = {A78},
          doi = {10.1051/0004-6361/201628536},
archivePrefix = {arXiv},
       eprint = {1603.04368},
 primaryClass = {astro-ph.CO},
       adsurl = {https://ui.adsabs.harvard.edu/abs/2017A&A...598A..78I}
}

@article{horvath2025long,
  title={A long period transient search method for the Murchison Widefield Array},
  author={Horv{\'a}th, Csan{\'a}d and Hurley-Walker, Natasha and McSweeney, Samuel J and Galvin, Timothy J and Morgan, John},
  journal={Publications of the Astronomical Society of Australia},
  pages={1--13},
  year={2025},
  publisher={Cambridge University Press}
}

@ARTICLE{2025ApJ...981..143M,
       author = {{Mcsweeney}, Samuel J. and {Moseley}, Jared and {Hurley-Walker}, Natasha and {Grover}, Garvit and {Horv{\'a}th}, Csan{\'a}d and {Galvin}, Timothy J. and {Meyers}, Bradley W. and {Tan}, Chia Min},
        title = "{Discovery of an RRAT-like Pulsar via Its Single Pulses in a Murchison Widefield Array Imaging Survey}",
      journal = {\apj},
         year = 2025,
        month = mar,
       volume = {981},
       number = {2},
          eid = {143},
        pages = {143},
          doi = {10.3847/1538-4357/adb27f},
archivePrefix = {arXiv},
       eprint = {2502.02130},
 primaryClass = {astro-ph.HE},
       adsurl = {https://ui.adsabs.harvard.edu/abs/2025ApJ...981..143M}
}

@ARTICLE{2023MNRAS.524.1291P,
       author = {{Padmanabh}, P.~V. and {Barr}, E.~D. and {Sridhar}, S.~S. and {Rugel}, M.~R. and {Damas-Segovia}, A. and {Jacob}, A.~M. and {Balakrishnan}, V. and {Berezina}, M. and {Bernadich}, M.~C. and {Brunthaler}, A. and {Champion}, D.~J. and {Freire}, P.~C.~C. and {Khan}, S. and {Kl{\"o}ckner}, H. -R. and {Kramer}, M. and {Ma}, Y.~K. and {Mao}, S.~A. and {Men}, Y.~P. and {Menten}, K.~M. and {Sengupta}, S. and {Venkatraman Krishnan}, V. and {Wucknitz}, O. and {Wyrowski}, F. and {Bezuidenhout}, M.~C. and {Buchner}, S. and {Burgay}, M. and {Chen}, W. and {Clark}, C.~J. and {K{\"u}nkel}, L. and {Nieder}, L. and {Stappers}, B. and {Legodi}, L.~S. and {Nyamai}, M.~M.},
        title = "{The MPIfR-MeerKAT Galactic Plane Survey - I. System set-up and early results}",
      journal = {\mnras},
         year = 2023,
        month = sep,
       volume = {524},
       number = {1},
        pages = {1291-1315},
          doi = {10.1093/mnras/stad1900},
archivePrefix = {arXiv},
       eprint = {2303.09231},
 primaryClass = {astro-ph.HE},
       adsurl = {https://ui.adsabs.harvard.edu/abs/2023MNRAS.524.1291P}
}

@software{adam_stewart_2024_14048598,
  author       = {Adam Stewart and
                  Serg and
                  Andrew O'Brien and
                  David Liptai and
                  Dougal Dobie and
                  Tom Mauch and
                  Shibli Saleheen and
                  Ella Xi Wang and
                  Joshua Pritchard and
                  Jazzy},
  title        = {askap-vast/vast-pipeline},
  month        = nov,
  year         = 2024,
  publisher    = {Zenodo},
  version      = {},
  doi          = {10.5281/zenodo.13927015},
  url          = {https://doi.org/10.5281/zenodo.13927015},
}

@software{adam_stewart_2025_15363128,
  author       = {Adam Stewart and
                  Dougal Dobie and
                  Andrew O'Brien and
                  David Kaplan},
  title        = {askap-vast/vast-tools},
  month        = may,
  year         = 2025,
  publisher    = {Zenodo},
  version      = {},
  doi          = {10.5281/zenodo.8365236},
  url          = {https://doi.org/10.5281/zenodo.8365236},
}

@ARTICLE{wsclean,
       author = {{Offringa}, A.~R. and {McKinley}, B. and {Hurley-Walker}, N. and {Briggs}, F.~H. and {Wayth}, R.~B. and {Kaplan}, D.~L. and {Bell}, M.~E. and {Feng}, L. and {Neben}, A.~R. and {Hughes}, J.~D. and {Rhee}, J. and {Murphy}, T. and {Bhat}, N.~D.~R. and {Bernardi}, G. and {Bowman}, J.~D. and {Cappallo}, R.~J. and {Corey}, B.~E. and {Deshpande}, A.~A. and {Emrich}, D. and {Ewall-Wice}, A. and {Gaensler}, B.~M. and {Goeke}, R. and {Greenhill}, L.~J. and {Hazelton}, B.~J. and {Hindson}, L. and {Johnston-Hollitt}, M. and {Jacobs}, D.~C. and {Kasper}, J.~C. and {Kratzenberg}, E. and {Lenc}, E. and {Lonsdale}, C.~J. and {Lynch}, M.~J. and {McWhirter}, S.~R. and {Mitchell}, D.~A. and {Morales}, M.~F. and {Morgan}, E. and {Kudryavtseva}, N. and {Oberoi}, D. and {Ord}, S.~M. and {Pindor}, B. and {Procopio}, P. and {Prabu}, T. and {Riding}, J. and {Roshi}, D.~A. and {Shankar}, N. Udaya and {Srivani}, K.~S. and {Subrahmanyan}, R. and {Tingay}, S.~J. and {Waterson}, M. and {Webster}, R.~L. and {Whitney}, A.~R. and {Williams}, A. and {Williams}, C.~L.},
        title = "{WSCLEAN: an implementation of a fast, generic wide-field imager for radio astronomy}",
      journal = {MNRAS},
         year = 2014,
        month = oct,
       volume = {444},
       number = {1},
        pages = {606-619},
          doi = {10.1093/mnras/stu1368},
archivePrefix = {arXiv},
       eprint = {1407.1943},
 primaryClass = {astro-ph.IM},
       adsurl = {https://ui.adsabs.harvard.edu/abs/2014MNRAS.444..606O}
}

@ARTICLE{pfb,
      title="{Africanus III. pfb-imaging -- a flexible radio interferometric imaging suite}", 
      author={Hertzog L. Bester and Jonathan S. Kenyon and Audrey Repetti and Simon J. Perkins and Oleg M. Smirnov and Tariq Blecher and Yassine Mhiri and Jakob Roth and Ian Heywood and Yves Wiaux and Benjamin V. Hugo},
      year={2025},
      eprint={2412.10073},
      archivePrefix={arXiv},
      primaryClass={astro-ph.IM},
      journal={Astronomy and Computing},
      volume={submitted}
}

@misc{cornwell2011askap,
  title={ASKAP science processing},
  author={Cornwell, T and Humphreys, B and Lenc, E and Voronkov, M and Whiting, M and Mitchell, D and Ord, S and Collins, D},
  journal={2016 [2018-09-13]. http://www. atnf. csiro. au/projects/askap/ASKAP-SW-0020. pdf},
  year={2011}
}

@article{Whiting_Humphreys_2012, title={Source-Finding for the Australian Square Kilometre Array Pathfinder}, volume={29}, DOI={10.1071/AS12028}, number={3}, journal={Publications of the Astronomical Society of Australia}, author={Whiting, M. and Humphreys, B.}, year={2012}, pages={371–381}}

@misc{lsst_brokers,
  author       = {{Vera C. Rubin Observatory}},
  title        = {Alerts and brokers},
  howpublished = {\url{https://rubinobservatory.org/for-scientists/data-products/alerts-and-brokers}},
  note         = {Accessed: 2025-11-14}
}

@MISC{2011ivoa.spec.0711S,
       author = {{Seaman}, Rob and {Williams}, Roy and {Allan}, Alasdair and {Barthelmy}, Scott and {Bloom}, Joshua and {Brewer}, John and {Denny}, Robert and {Fitzpatrick}, Mike and {Graham}, Matthew and {Gray}, Norman and {Hessman}, Frederic and {Marka}, Szabolcs and {Rots}, Arnold and {Vestrand}, Tom and {Wozniak}, Przemyslaw},
        title = "{Sky Event Reporting Metadata Version 2.0}",
 howpublished = {IVOA Recommendation 11 July 2011},
         year = 2011,
        month = jul,
        pages = {711},
          doi = {10.5479/ADS/bib/2011ivoa.spec.0711S},
archivePrefix = {arXiv},
       eprint = {1110.0523},
 primaryClass = {astro-ph.IM},
       adsurl = {https://ui.adsabs.harvard.edu/abs/2011ivoa.spec.0711S}
}

\end{document}